\documentclass[11pt]{article}
\usepackage{graphicx}
\usepackage{latexsym,graphicx}
\usepackage{float}
\usepackage{amsmath}
\usepackage{amsfonts}
\usepackage{amssymb}
\usepackage{epstopdf}
\usepackage{color}
\DeclareGraphicsExtensions{.eps}
\usepackage[left=2.5cm,top=2.5cm,right=2.5cm,bottom=2.5cm]{geometry}

\catcode `\@=11
\catcode `\@=12
\begin{document}
	\begin{center}
		\large{\bf{Transit String Dark Energy Models in $f(Q)$ Gravity}} \\
		\vspace{5mm}
		\normalsize{ Dinesh Chandra Maurya $^1$, Archana Dixit $^2$, Anirudh Pradhan $^3$ }\\
		\vspace{5mm}
			
		\normalsize{$^{1}$Department of Mathematics, Faculty of Sciences, IASE (Deemed to be University), Sardarshahar-331 403 (Churu), Rajsthan, India.}\\
	
		\vspace{2mm}
		\normalsize{$^{2}$Department of Mathematics, Institute of Applied Sciences and Humanities, GLA University, Mathura-281 406, Uttar Pradesh, India}\\
	
			\vspace{2mm}
			\normalsize{$^{3}$ Centre for Cosmology, Astrophysics and Space Science (CCASS), GLA University,\\
			Mathura-281 406, Uttar Pradesh, India}\\
			\vspace{2mm}
		$^1$E-mail:dcmaurya563@gmail.com \\
			\vspace{2mm}
		$^2$E-mail:archana.dixit@gla.ac.in \\
			\vspace{2mm}
		$^3$E-mail:pradhan.anirudh@gmail.com\\

	\end{center}
	\vspace{5mm}
	\begin{abstract}
      In this paper, we have investigated an anisotropic cosmological model in $f(Q)$ gravity with string fluid in LRS Bianchi type-I universe. We have considered the arbitrary function $f(Q)=Q+\alpha\sqrt{~Q}+2\Lambda$ where $\alpha$ is model free parameter and $\Lambda$ is the cosmological constant. We have established a relationship between matter energy density parameter $\Omega_{m}$ and dark energy density parameter $\Omega_{\Lambda}$ through Hubble function using using constant equation of state parameter $\omega$. We have made observational constraint on the model using $\chi^2$-test with observed Hubble datasets $H(z)$ and SNe Ia datasets, and obtained the best fit values of cosmological parameters. We have used these best fit values in the result and discussion. We have discussed our result with cosmographic coefficients and found a transit phase dark energy model. Also, we analyzed the Om diagnostic function for anisotropic universe and found that our model is quintessence dark energy model.
	\end{abstract}
	\smallskip
	\vspace{5mm}
	
	\section{Introduction}
The observational studies in the last decade of the twentieth century confirm and support the acceleration in expansion of the universe \cite{ref1}-\cite{ref7}. But, the theoretical development of the accelerating cosmological models have many real issues. Although, the theory of general relativity (GR) is still the most successful fundamental theory of gravity to describe the large-scale evolution of the universe. First time, Einstein has applied GR to cosmology and found an universe model which is either collapsing or expanding. At that time, the universe was supposed to be static; there was no evidence of an accelerating or expanding universe in literature. So, Einstein added a constant in his field equation to obtain a static universe which is called as "Cosmological Constant $\Lambda$" but he removed this cosmological constant from his field equation after the Hubble discovery of the expanding universe in 1929. Einstein again added this cosmological constant to his field equation after the discovery of accelerating expanding universe in 1998 but this time, the cosmological constant $\Lambda$ was added to explain the dark energy problem which is supposed to cause of acceleration in expanding universe and this model is called as $\Lambda$CDM model which well fitted with observational datasets and well explain the late-time accelerating universe. Although, this $\Lambda$-term is well fitted to explain the dark energy problem, but it has two significant problem, one is its origin in equation and second is its fine-tuning problems related to the vacuum energy scale \cite{ref8}-\cite{ref10}. After that many cosmologists have tried to modify Einstein field equations and some have tried to present alternate theories of gravity, to explain this dark energy problem in the universe.\\

The string cosmological model currently plays a significant role in explaining the universe's early stages since it unifies descriptions of the early universe's fundamental nature and structure. All matter and forces are combined into a single theoretical framework by the theory of strings, which describes the fundamental level of the cosmos in terms of vibrating strings rather than particles. In general, people thought that cosmic strings were a type of one-dimensional topological defect that may have developed as a result of the early universe's broken symmetry during a phase transition as a result of the universe's temperature cooling below a certain threshold in the early stages of the universe after the Big Bang.\\

After that, the study of cosmological models combined with cosmic strings in general relativity (GR) and in alternative theories of gravitation has occasionally motivated many cosmologists. The significance of cosmic strings in general relativity and their gravitational effects have been thoroughly examined by Kibble in 1976, 1980, and 1982, as well as by Kibble and Turok \cite{ref11}. Various elements of cosmic strings combined with perfect fluid and electromagnetic field in general relativity have been studied by several writers \cite{ref12,ref13,ref14,ref15}. Letelier \cite{ref16} created a cosmological model using cosmic strings within the context of the standard GR that is the basis for many other cosmological models, including those of the Bianchi-I and Kantowski-Sach types. Krori et al. \cite{ref17} examined spatially homogeneous models of the Bianchi types $II, VI 0, VII$, and $IX$ in the presence of strings. By adding the source matter stress energy tensor for a perfect dust together with cosmic strings, several cosmological models can be generalised to null strings and to perfect fluid strings. From the vanishing divergence equation of the stress energy tensor \cite{ref14}, one can obtain the equations of motion of strings and the conservation laws of dust. In the context of general theory of relativity, numerous scholars \cite{ref18} have examined the string cosmological model in the magnetic field era. We have investigated the string cosmological models in \cite{ref19,ref20}.\\


The simplest generalization of GR is the so-called $f(R)$-gravity proposed in \cite{ref21,ref22} which is obtained by replacing the ricci-scalar $R$ by an arbitrary function of $R$ in the Hilbert-Einstein action. The modified $f(R)$ gravity is well-known for its successful explanation of cosmic acceleration, and also it can reproduce the whole evolution history of the universe, as well as the evolution of cosmological constant $\Lambda$-term \cite{ref23,ref24,ref25}. In order to obtain this cosmic acceleration through modified theories of gravity, there are several techniques adopted in the literature.\\

The amazing thing about this supposition is that it assumes that matter fields have no impact whatsoever on gravitational interaction. To express the affine features of a manifold, however, the curvature is not the only geometric object that may be used (see references \cite{ref26}-\cite{ref29}). In actuality, torsion and non-metricity are the other two fundamental items connected to the connection of a metric space in addition to curvature. Torsion and non-metricity disappear in Einstein's conventional General Relativity (GR). If we accept the equivalence principle's claim that gravity has a geometrical nature, it is important to investigate the different ways that gravity can have an analogous geometry. If one takes into account a flat spacetime with a metric but asymmetric connectivity, GR is represented equivalently. Torsion is given complete control over gravity in this teleparallel formulation. On an identically flat spacetime without torsion, a third equivalent and more straightforward model of GR can be built, in which gravity is this time attributed to non-metricity. Thus, the Einstein-Hilbert action $\int \sqrt{-g} R(g)$, the teleparallel equivalent of GR $\int \sqrt{-g}T$ \cite{ref30} and coincident GR $\int \sqrt{-g}Q$ \cite{ref31}, both of which rest on a symmetric teleparallel geometry \cite{ref32}, can all be used to describe the same underlying physical theory, GR. You can read more about the geometrical trinity of GR and its compact presentation in \cite{ref29}.

The underlying underpinning of these geometrical interpretations offers modified gravity a potential path forward. Once the appropriate scalar values are, for example, promoted to arbitrary functions thereof, the analogous formulations of GR with curvature,  non-metricity and torsion, represent different possible starting points to modified gravity theories. We will concentrate on the less researched example of $f(Q)$ theories, which was introduced for the first time in \cite{ref31}, because it is less investigated than modified gravity theories based on $f(R)$ \cite{ref33,ref34,ref35} and $f(T)$ \cite{ref36,ref37,ref38}. Recently J. Baltran et al. \cite{ref39} have investigated various features of cosmology in $f(Q)$ geometry, although $f(Q)$ cosmography with energy
conditions can be found in \cite{ref40, ref41} whereas cosmological clarifications and evolution index of matter perturbations have been examined
for a polynomial functional form of $f(Q)$ \cite{ref42}. Harko et al. investigated the connection matter in $f(Q)$ gravity by supposing a power-law
function \cite{ref43}. Recently, ``we have studied the transit phase universe in $f(Q, T)$ gravity \cite{ref44}, dark energy nature of bulk viscosity in an isotropic universe with $f(Q)$ gravity in \cite{ref45,ref46,ref47}". Recently, Koussour et al. \cite{ref48} have studied anisotropic $f(Q)$ gravity with bulk viscosity.\\
As a result of the foregoing arguments, we shall examine in this paper, cosmological aspects of $f(Q)$-gravity that are not previously covered in the literature using string fluid. Here, we will select a particular version of $f(Q)=Q+\alpha\sqrt{Q}+2\Lambda$ that results after integrating the field equation $Qf_{Q}-\frac{f}{2}=\rho$ using the generalised $\Lambda$CDM field equation $3H^{2}=\rho+\Lambda$ in a flat FLRW spacetime. Here, $\alpha$ is called as model free parameter. Recently Capozziello and D'Agostino \cite{ref49} have obtained $f(Q)=Q+\beta$, $\beta>0$ for $\Lambda$CDM model by using model-independent reconstruction method that can be find from $f(Q)=Q+\alpha\sqrt{Q}+2\Lambda$ by putting $\alpha=0$. And in this article, we'll investigate this particular type of $f(Q)$ gravity in an anisotropic universe with a string dust-fluid source.\\
The present study is organised as follows: Sect.-1 is introductory, Sect.-2 contains basic concepts and formulation of field equations in $f(Q)$-gravity, the solutions of the field equations are given in sect.-3 and in sect.-4, we have made some observational constraints on the model. In sect.-5, we have obtained result and discussion. In last sect.-6 conclusions are done.
\section{Field Equations in $f(Q)$ Gravity}
Consider a Palatini-style framework in which connections and metrics are treated on equal footing, such that they are independent objects and their relationships are imposed only by field equations. In this model, the space-time manifold has a metric structure that comes from the metric $g_{\mu\nu}$, while affine connections ${\Gamma^{\alpha}}_{\mu\nu}$ are returns the affine structure that reveals, how tensors are transported,that define covariant derivatives.
In the group of theories that this work looks at, the non-metricity tensor is the most important thing and defined as $Q_{\alpha\mu\nu}=\nabla_{\alpha}g_{\mu\nu}$. Compatibility of metrics reveals connection failures. We can derive the deformation from the non-metric tensor $Q_{\alpha\mu\nu}$ as

\begin{equation}\label{eq1}
  {L^{\alpha}}_{\mu\nu}=\frac{1}{2}{Q^{\alpha}}_{\mu\nu}-{Q_{(\mu\nu)}}^{\alpha},
\end{equation}
and it estimates the advancements of the symmetric part of the full connection from the Levi-Civita connection. It will also
be convenient to introduce the non-metricity conjugate defined as
\begin{equation}\label{eq2}
  {P^{\alpha}}_{\mu\nu}=-\frac{1}{2}{L^{\alpha}}_{\mu\nu}+\frac{1}{4}(Q^{\alpha}-\tilde{Q^{\alpha}})g_{\mu\nu}-\frac{1}{4}\delta^{\alpha}_{(\mu Q_{\nu})}
\end{equation}
where ``$Q_{\alpha}=g^{\mu\nu}Q_{\alpha\mu\nu}$ and $\tilde{Q_{\alpha}}=g^{\mu\nu}Q_{\mu\alpha\nu}$ are two independent traces of the non-metricity tensor". Now, we will define the non-metricity scalar $Q$
that is the most important part of this work, as
\begin{equation}\label{eq3}
  Q=-Q_{\alpha\mu\nu}P^{\alpha\mu\nu}.
\end{equation}
One can observe why we say $P^{\alpha\mu\nu}$ the non-metricity conjugate since it derived as
\begin{equation}\label{eq4}
  P^{\alpha\mu\nu}=-\frac{1}{2}\frac{\partial{Q}}{\partial{Q_{\alpha\mu\nu}}}
\end{equation}
Thus, we can write the action for our model in the form of an arbitrary function of non-metricity scalar $Q$ as below

\begin{equation}\label{eq5}
  S=\int{d^{4}x~\sqrt{-g}\left[-\frac{1}{2}f(Q)+L_{m}\right]}
\end{equation}
where the matter Lagrangian is represented by $L m$. The reason for the specific selection of the non-metricity scalar and the aforementioned action is that, for the choice $f=Q/8\pi G$ \cite{ref31}, we recover the so-called ``Symmetric Teleparallel Equivalent of GR (classically, up to a boundary term").\\

Now, taking variation of action (\ref{eq5}), the metric field equations are obtained as
\begin{equation}\label{eq6}
  \frac{2}{\sqrt{-g}}\nabla_{\alpha}({\sqrt{-g}f_{Q}{P^{\alpha}}_{\mu\nu}})+\frac{1}{2}g_{\mu\nu}f+f_{Q}(P_{\mu\alpha\beta}{Q_{\nu}}^{\alpha\beta}-2Q_{\alpha\beta\mu}{P^{\alpha\beta}}_{\nu})=T_{\mu\nu}
\end{equation}
where $f_{Q}=\partial{f}/\partial{Q}$. This takes on an even somewhat more compact shape by rising one index.,
\begin{equation}\label{eq7}
  \frac{2}{\sqrt{-g}}\nabla_{\alpha}({\sqrt{-g}f_{Q}{P^{\alpha\mu}}_{\nu}})+\frac{1}{2}\delta^{\mu}_{\nu}f+f_{Q}P^{\mu\alpha\beta}Q_{\nu\alpha\beta}=T^{\mu}_{\nu}
\end{equation}

By observing that the connection's variation with respect to $\xi^{\alpha}$ is comparable to conducting a diffeomorphism, the connection equation of motion may be easily calculated as $\delta_{\xi}{\Gamma^{\alpha}}_{\mu\beta}=-L_{\xi}{\Gamma^{\alpha}}_{\mu\beta}=-\nabla_{\mu}\nabla_{\beta}\xi^{\alpha}$, the connection is torsion-free and in the location where we have utilised it. Therefore, the connection field equations read as follows in the absence of hypermomentum.
\begin{equation}\label{eq8}
  \nabla_{\mu}\nabla_{\nu}(\sqrt{-g}f_{Q}{P^{\mu\nu}}_{\alpha})=0
\end{equation}
One may confirm that ``$D_{\mu}{T^{\mu}}_{\nu}=0$, where $D_{\mu}$ is the metric-covariant derivative \cite{ref50}", exists from the metric and connection equations, as it should due to diffeomorphism (Diff) invariance. The divergence of the energy-momentum tensor and the hypermomentum would be related in the most general case with a nontrivial hypermomentum, as shown in reference \cite{ref43}.\\

Recently, we built dark energy models in a flat FLRW universe with bulk viscosity in $f(Q)$-gravity (see references \cite{ref45, ref46, ref47}), and in this study, we will derive a string-fluid dark energy model with any function $f(Q)=Q+\alpha \sqrt{Q}+\Lambda$ in LRS Bianchi Type-I space-time metric:

\begin{equation}\label{eq9}
  ds^{2}=A(t)^{2}dx^{2}+B(t)^{2}(dy^{2}+dz^{2})-dt^{2}
\end{equation}
Here, A(t) and B(t) are functions of cosmic time $t$ and are referred to as metric coefficients. The non-metricity scalar $Q$ for the line element (\ref{eq9}) is calculated as
\begin{equation}\label{eq10}
  Q=4\frac{\dot{A}}{A}\frac{\dot{B}}{B}+2\left(\frac{\dot{B}}{B}\right)^{2}
\end{equation}
For a string-fluid, the stress-energy-momentum tensor is regarded as
 \begin{equation}\label{eq11}
 T_{\mu\nu}=(p+\rho)u_{\mu}u_{\nu}+p g_{\mu\nu}-\lambda x_{\mu} x_{\nu},
 \end{equation}
 where $u^{\mu}=(0,0,0,-1)$ is the four-velocity vector in the co-moving coordinate system satisfying $u_{\mu} u^{\mu}=-x_{\mu} x^{\mu}=-1, u_{\mu} x^{\mu}=0, x^{\mu}$ is the direction of the string and $u^{\mu}\nabla_{\nu} u_{\mu}=0$. Here, $p$ stands for the fluid's pressure, $\lambda$ represents the string tension density, and $\rho$ represents the energy density for the cloud of threads with attached particles. The definition of $\rho_{p}$, the particle density, is
 \begin{equation}\label{eq12}
 \rho=\rho_p+\lambda,
 \end{equation}
where $\lambda$ could be either positive or negative. \cite{ref51}.\\
The significance of cosmic strings in general relativity and their gravitational effects have been thoroughly examined by Kibble in 1976, 1980, and 1982, as well as by Kibble and Turok \cite{ref11}. Various elements of cosmic strings combined with perfect fluid and electromagnetic field in general relativity have been studied by several writers \cite{ref12,ref13,ref14,ref15}. Letelier \cite{ref16} created a cosmological model using cosmic strings within the context of the standard GR that is the basis for many other cosmological models, including those of the Bianchi-I and Kantowski-Sach types.\\
In a co-moving coordinate system, we can obtain the following field equations from Eqs.~(\ref{eq7}), (\ref{eq9}) and (\ref{eq11}) as given below
\begin{equation}\label{eq13}
  f_{Q}\left[4\frac{\dot{A}}{A}.\frac{\dot{B}}{B}+2\left(\frac{\dot{B}}{B}\right)^{2}\right]-\frac{f}{2}=\rho
\end{equation}
\begin{equation}\label{eq14}
  2f_{Q}\left[\frac{\dot{A}}{A}.\frac{\dot{B}}{B}+ \left(\frac{\dot{B}}{B}\right)^{2}+\frac{\ddot{B}}{B}+\right]+
  2\frac{\dot{B}}{B}\dot{Q}f_{QQ} -\frac{f}{2}=-p+\lambda
\end{equation}
\begin{equation}\label{eq15}
  f_{Q}\left[3\frac{\dot{A}}{A}.\frac{\dot{B}}{B}+ \left(\frac{\dot{B}}{B}\right)^{2}+\frac{\ddot{A}}{A}+\frac{\ddot{B}}{B}+
 \right]+\left(\frac{\dot{A}}{A}+\frac{\dot{B}}{B}\right)\dot{Q}f_{QQ}-\frac{f}{2}=-p
\end{equation}
The matter conservation law is given by
\begin{equation}\label{eq16}
  \dot{\rho}+3H(\rho+p)-\lambda\frac{\dot{A}}{A}=0
\end{equation}
the derivative with respect to time $t$ is represented here by the over dot symbol $(^{.})$.\\

Now, we define the various cosmological parameters viz. volume scale-factor $V(t)=a(t)^{3}=AB^{2}$ where $a(t)$ is the average scale-factor, the deceleration parameter $q(t)=-\frac{a\ddot{a}}{\dot{a}^{2}}$ that shows the phase of the expanding universe. The mean Hubble parameter is defined as $H=\frac{1}{3}(H_{x}+H_{y}+H_{z})$, with the directional Hubble parameter $H_{x}=\frac{\dot{A}}{A},~H_{y}=H_{z}=\frac{\dot{B}}{B}$.\\

Now, we define some more geometrical parameters $\theta,~\sigma^{2},~\Delta$ called as scalar expansion, shear scalar and the
mean anisotropy parameter respectively and are given by
\begin{equation}\label{eq17}
  \theta(t)=\frac{\dot{A}}{A}+2.\frac{\dot{B}}{B}
\end{equation}
\begin{equation}\label{eq18}
  \sigma^{2}(t)=\frac{1}{3}\left(\frac{\dot{A}}{A}-\frac{\dot{B}}{B}\right)^{2}
\end{equation}
\begin{equation}\label{eq19}
  \Delta=\frac{1}{3}\sum_{i=x}^{z}\left(\frac{H_{i}-H}{H}\right)^{2}
\end{equation}
where $H_{i}, i=x, y, z$, are the directional Hubble parameters .
\section{Solution of the field equations}
Here, we have three linearly independent equations (\ref{eq13}), (\ref{eq14}) \& (\ref{eq15}) in five unknowns $A,~B,~\rho,~p$ and $\lambda$. To find exact solutions of the field equations, we have required at least two constraints on these parameters. For a spatially homogeneous metric, the normal congruence to homogeneous expansion implies that the expansion scalar $\theta$ is proportional to shear scalar $\sigma$ . ie. $\sigma/\theta=k$ where $k$ is constant. This condition leads to \cite{ref52}
\begin{equation}\label{eq20}
A=B^{k}
\end{equation}
where $k$ is a constant.\\
Secondly, we define the equation of state (EoS) for the matter as $p=\omega\rho$ and $\lambda=\omega_{\lambda}\rho$ with $\omega$ as EoS parameter for perfect fluid matter and $\omega_{\lambda}$ is EoS parameter for string fluid.\\
Now, using Eq.~(\ref{eq20}) in volume-scale factor, we get
\begin{equation}\label{eq21}
  A=a^{\frac{3k}{k+2}},~~B=a^{\frac{3}{k+2}}
\end{equation}
Using Eq.~(\ref{eq21}) in (\ref{eq16}) and integrating, we get the energy density $\rho$ as
\begin{equation}\label{eq22}
  \rho=\rho_{0}\left(\frac{a_{0}}{a}\right)^{3(1+\omega)-\frac{3k\omega_{\lambda}}{k+2}}
\end{equation}
and using redshift relation $\frac{a_{0}}{a}=1+z$, we have
\begin{equation}\label{eq23}
  \rho=\rho_{0}(1+z)^{3(1+\omega)-\frac{3k\omega_{\lambda}}{k+2}}
\end{equation}
Now, we consider the function $f(Q)$ as

\begin{equation}\label{eq24}
  f(Q)=Q+\alpha\sqrt{~Q}+2\Lambda
\end{equation}
where $\alpha$ is model free parameter and $\Lambda$ is the cosmological constant. Recently Capozziello and D'Agostino \cite{ref49} have obtained $f(Q)=Q+\beta$, $\beta>0$ for $\Lambda$CDM model by using model-independent reconstruction method that can be find from $f(Q)=Q+\alpha\sqrt{Q}+2\Lambda$ by putting $\alpha=0$.\\
Now, using Eqs.~(\ref{eq21}), (\ref{eq23}) and (\ref{eq24}) in (\ref{eq13}), we get
\begin{equation}\label{eq25}
  \Omega_{m}+\Omega_{\Lambda}=\frac{3(2k+1)}{(k+2)^{2}}
\end{equation}
where $\Omega_{m}=\frac{\rho}{3H^{2}}$ and $\Omega_{\Lambda}=\frac{\Lambda}{3H^{2}}$, called as matter and dark energy density parameter respectively. Using Eq.~(\ref{eq23}) in (\ref{eq25}) and taking dust-fluid $\omega=0$, we have
\begin{equation}\label{eq26}
  H(t)=H_{0}\frac{k+2}{\sqrt{3(2k+1)}}\sqrt{\Omega_{m0}\left(\frac{a_{0}}{a}\right)^{3-\frac{3k\omega_{\lambda}}{k+2}}+\Omega_{\Lambda0}}
\end{equation}
or
\begin{equation}\label{eq27}
  H(z)=H_{0}\frac{k+2}{\sqrt{3(2k+1)}}\sqrt{\Omega_{m0}(1+z)^{3-\frac{3k\omega_{\lambda}}{k+2}}+\Omega_{\Lambda0}}
\end{equation}
Now, subtracting Eq.~(\ref{eq14}) from (\ref{eq15}) and using Eqs.~(\ref{eq21}) - (\ref{eq27}), we get the string tension density as
\begin{multline}\label{eq28}
 \lambda=\frac{3H_{0}^{2}\Omega_{\Lambda0}(k-1)(k+2)}{2k+1}+\frac{3H_{0}^{2}\Omega_{m0}(k-1)[(k+2)-k\omega_{\lambda}]}{2(2k+1)}(1+z)^{3-\frac{3k\omega_{\lambda}}{k+2}}+\frac{3\alpha H_{0}(k-1)(k+2)}{2\sqrt{6}(2k+1)}\times\\
 \sqrt{\Omega_{m0}(1+z)^{3-\frac{3k\omega_{\lambda}}{k+2}}+\Omega_{\Lambda0}}
\end{multline}
The deceleration parameter $q$ is obtained as
\begin{multline}\label{eq29}
  q(z)=\frac{2(k^{2}-k+3)}{k+3}-\frac{3\omega_{\lambda}(2k+1)}{(k+2)(k+3)}+\frac{\alpha\sqrt{3}(k^{2}-3k+2)}{2\sqrt{2}H_{0}(k+2)(k+3)}\frac{1}{\sqrt{\Omega_{m0}(1+z)^{3-\frac{3k\omega_{\lambda}}{k+2}}+\Omega_{\Lambda0}}}\\
  -\frac{6(2-\omega_{\lambda})(2k+1)}{(k+2)(k+3)}\frac{\Omega_{\Lambda0}}{[\Omega_{m0}(1+z)^{3-\frac{3k\omega_{\lambda}}{k+2}}+\Omega_{\Lambda0}]}
\end{multline}
\section{Observational Constraints}
\subsection{Hubble Parameter}
In this part of the paper, we have obtain best fit curve of Hubble function as mentioned in Eq. (\ref{eq27}), using $46$ Hubble constant observed values of $H(z)$ with error over redshift $z$ which is obtained in \cite{ref53}-\cite{ref68} using differential age (DA) method time to time (see Table 1).
 Here, we have used $\chi^{2}$-test formula as given below:
\begin{equation}\label{eq30}
  \chi^{2}=\sum_{i=1}^{i=N}\frac{[O_{i}-E_{i}]^{2}}{\sigma_{i}^{2}}
\end{equation}
where $N$ denotes the number of data, $O_{i},~E_{i}$ represent the observed and estimated datasets respectively and $\sigma_{i}$ denotes standard deviations.
\begin{figure}[H]
\centering
	\includegraphics[width=10cm,height=6cm,angle=0]{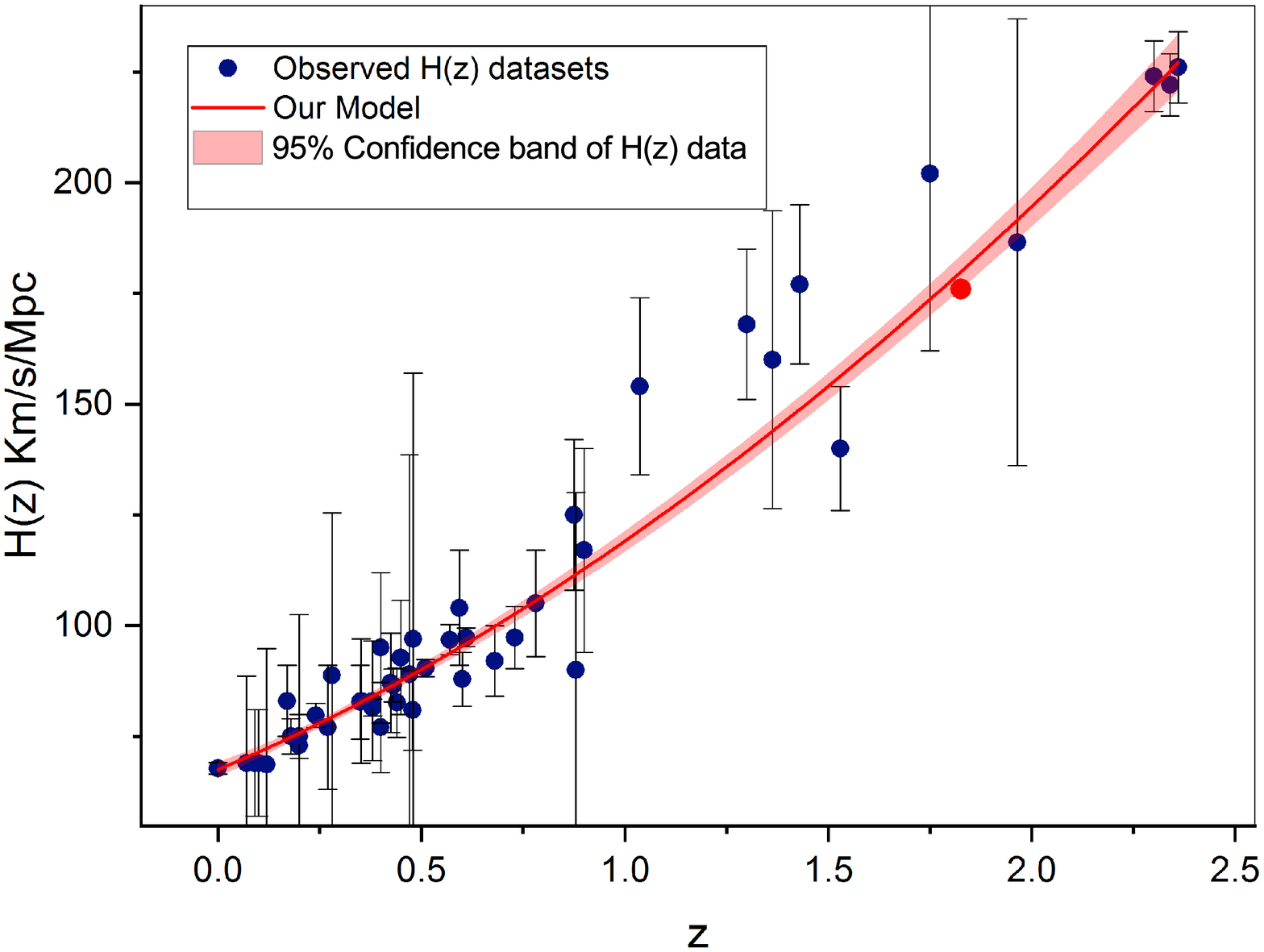}
  	\caption{Best fit curve of Hubble parameter $H(z)$ versus redshift $z$ with observed $H(z)$ data sets.}
\end{figure}

 \begin{table}[H]
  \centering
 {\tiny \begin{tabular}{|c|c|c|c|c|c|c|c|c|c|c|c|}
     \hline
			``S.No. & $z$   & $H(z)$  & $\sigma_{H}$  & Reference           & S.No. & $z$   & $H(z)$  & $\sigma_{H}$  & Reference \\
			\hline
			1  & $0$      & $67.77$ & $1.30$   & \cite{ref53}         & 24  & $0.4783$  & $80.9$  & $9$     & \cite{ref67}  \\
			2  & $0.07$   & $69$    & $19.6$   & \cite{ref54}         & 25  & $0.48$    & $97$    & $60$    & \cite{ref55}  \\
			3  & $0.09$   & $69$    & $12$     & \cite{ref66}         & 26  & $0.51$    & $90.4$  & $1.9$   & \cite{ref57}  \\
			4  & $0.10$   & $69$    & $12$     & \cite{ref55}         & 27  & $0.57$    & $96.8$  & $3.4$   & \cite{ref68}  \\
			5  & $0.12$   & $68.6$  & $26.2$   & \cite{ref54}         & 28  & $0.593$   & $104$   & $13$    & \cite{ref65}  \\
			6  & $0.17$   & $83$    & $8$      & \cite{ref55}         & 29  & $0.60$    & $87.9$  & $6.1$   & \cite{ref59}  \\
			7  & $0.179$  & $75$    & $4$      & \cite{ref65}         & 30  & $0.61$    & $97.3$  & $2.1$   & \cite{ref57}  \\
			8  & $0.1993$ & $75$    & $5$      & \cite{ref65}         & 31  & $0.68$    & $92$    & $8$     & \cite{ref65}  \\
			9  & $0.2$    & $72.9$  & $29.6$   & \cite{ref54}         & 32  & $0.73$    & $97.3$  & $7$     & \cite{ref59}  \\
		   10  & $0.24$   & $79.7$  & $2.7$    & \cite{ref56}         & 33  & $0.781$   & $105$   & $12$    & \cite{ref65}  \\
		   11  & $0.27$   & $77$    & $14$     & \cite{ref55}         & 34  & $0.875$   & $125$   & $17$    & \cite{ref65}  \\
		   12  & $0.28$   & $88.8$  & $36.6$   & \cite{ref54}         & 35  & $0.88$    & $90$    & $40$    & \cite{ref55}  \\
		   13  & $0.35$   & $82.7$  & $8.4$    & \cite{ref58}         & 36  & $0.9$     & $117$   & $23$    & \cite{ref55}  \\
		   14  & $0.352$  & $83$    & $14$     & \cite{ref65}         & 37  & $1.037$   & $154$   & $20$    & \cite{ref56}  \\
		   15  & $0.38$   & $81.5$  & $1.9$    & \cite{ref57}         & 38  & $1.3$     & $168$   & $17$    & \cite{ref55}  \\
		   16  & $0.3802$ & $83$    & $13.5$   & \cite{ref58}         & 39  & $1.363$   & $160$   & $33.6$  & \cite{ref61}  \\
           17  & $0.4$    & $95$    & $17$     & \cite{ref66}         & 40  & $1.43$    & $177$   & $18$    & \cite{ref55}  \\
           18  & $0.004$  & $77$    & $10.2$   & \cite{ref67}         & 41  & $1.53$    & $140$   & $14$    & \cite{ref55}  \\
           19  & $0.4247$ & $87.1$  & $11.2$   & \cite{ref67}         & 42  & $1.75$    & $202$   & $40$    & \cite{ref61}  \\
           20  & $0.43$   & $86.5$  & $3.7$    & \cite{ref56}         & 43  & $1.965$   & $186.5$ & $50.4$  & \cite{ref56}  \\
	       21  & $0.44$   & $82.6$  & $7.8$    & \cite{ref59}         & 44  & $2.3$     & $224$   & $8$     & \cite{ref64}  \\
           22  & $0.44497$& $92.8$  & $12.9$   & \cite{ref67}         & 45  & $2.34$    & $222$   & $7$     & \cite{ref62}  \\
           23  & $0.47$   & $89$    & $49.6$ "  & \cite{ref60}         & 46  & $2.36$    & $226$   & $8$"     & \cite{ref63}  \\
     \hline
   \end{tabular}}
  \caption{Hubble's constant table.}\label{T1}
\end{table}
On the basis of the minimum $\chi^{2}$, the estimated present values of various parameters are  shown in the following tables.
\begin{table}[H]
  \centering
  \begin{tabular}{|c|c|}
     \hline

			Parameter            & Values         \\
			\hline
		
            $\Omega_{m0}$        & $0.41392\pm0.01866$   \\

            $H_{0}$              & $68.60951\pm1.65678$   \\

            $k$                  & $0.30652\pm0.01114$   \\

            $\omega_{\lambda}$   & $0.9417$   \\

            $\chi^{2}$           & $0.0617973$   \\
     \hline
   \end{tabular}
  \caption{The best fit values of $\Omega_{m0},~H_{0},~\omega_{\lambda}$ in $\chi^{2}$-test of $H(z)$ with observed $H(z)$ data with $95\%$ confidence level of bounds. (see Figure 1 ).}\label{T2}
\end{table}
Eq.~(\ref{eq27}) shows the expression for the Hubble function and its best fit curve is shown in Figure 1 for the best fit values of cosmological parameters $\Omega_{m0},~H_{0},~k,~\omega_{\lambda}$ as mentioned in Table 2. We have estimated the corresponding present value of dark energy density parameter $\Omega_{\Lambda0}=0.4957$ and current value of matter and dark energy densities are measured as $\rho_{0}=6.1378\times10^{-36}~gm/cm^{3}$ and $\Lambda=7.3505\times10^{-36}~gm/cm^{3}$.
 \subsection{Luminosity Distance}
The redshift-luminosity distance connection is a crucial observational tool for understanding the universe's evolution. The expression for the luminosity distance ($D_{L}$) is calculated in terms of redshift because the universe's expansion causes the light emitted by distant luminous bodies to undergo a redshift. It is provided as
\begin{equation}\label{eq31}
	D_{L}=a_{0} r (1+z).
\end{equation}
where $r$ is the source's radial coordinate and is determined by

\begin{equation}\label{eq32}
	r  =  \int^r_{0}dr = \int^t_{0}\frac{cdt}{a(t)} = \frac{1}{a_{0}H_{0}}\int^z_0\frac{cdz}{h(z)}
\end{equation}
where we have used $ dt=dz/\dot{z}, \dot{z}=-H(1+z)~\&~ h(z)=\frac{H}{H_0}.$

Therefore  the luminosity distance is obtained as:

\begin{equation}\label{33}
	D_{L}=\frac{c(1+z)}{H_{0}}\int^z_0\frac{dz}{h(z)}
\end{equation}

\subsection{ Distance modulus $\mu(z)$ and Apparent Magnitude $m(z)$}
The distance modulus $\mu$ is derived as \cite{ref71}:

\begin{eqnarray*}
	\mu & = &  m_{b}-M     \\
	& = &  5log_{10}\left(\frac{D_L}{Mpc}\right)+25 \\
	& = & 25+  5log_{10}\left[\frac{c(1+z)}{H_0} \int^z_0\frac{dz}{h(z)}\right].
\end{eqnarray*}
\begin{equation}\label{eq34}
\end{equation}
The absolute magnitude $M$ of a supernova \cite{ref72,ref73} is obtained as :
\begin{center}
	\begin{equation}\label{eq35}
		M=16.08-25+5log_{10}(H_{0}/.026c)
	\end{equation}
	\par\end{center}
Eqs. (\ref{eq34}) and (\ref{eq35}) produce the following expression for the  apparent magnitude $m(z)$

\begin{equation}\label{eq36}
	m(z)=16.08+ 5log_{10}\left[\frac{1+z}{.026} \int^z_0\frac{dz}{h(z)}\right].
\end{equation}
We  adopt the most recent compilation of supernovae pantheon samples which consist
of 715 SN Ia data in the range of ($ 0.01 \le  z \le 1.414$ ) \cite{ref74,ref75}. For constraining various parameter of the model, we have used the following $\chi^2$ formula:
\begin{equation}\nonumber
  \chi^{2}=\sum_{i=1}^{i=N}\frac{[O_{i}-E_{i}]^{2}}{\sigma_{i}^{2}}
\end{equation}
where $N$ denotes the number of data, $O_{i},~E_{i}$ represent the observed and estimated datasets respectively and $\sigma_{i}$ denotes standard deviations.\\
On the basis of the minimum $\chi^{2}$, the estimated present values of various parameters are  shown in the following tables.
\begin{table}[H]
  \centering
  \begin{tabular}{|c|c|}
     \hline

			Parameter            & Values         \\
			\hline
		
            $\Omega_{m0}$        & $0.43061\pm0.02755$   \\

            $k$                  & $0.30242\pm0.0072$   \\

            $\omega_{\lambda}$   & $0.86073$   \\

            $\chi^{2}$           & $0.072955$   \\
     \hline
   \end{tabular}
  \caption{The best fit values of $\Omega_{m0},~k,~\omega_{\lambda}$ in $\chi^{2}$-test of apparent magnitude $m(z)$ with observed $m(z)$ from SNe Ia data with $95\%$ confidence level of bounds. (see Figure 2a ).}\label{T3}
\end{table}
\begin{figure}[H]
	a.\includegraphics[width=9cm,height=6cm,angle=0]{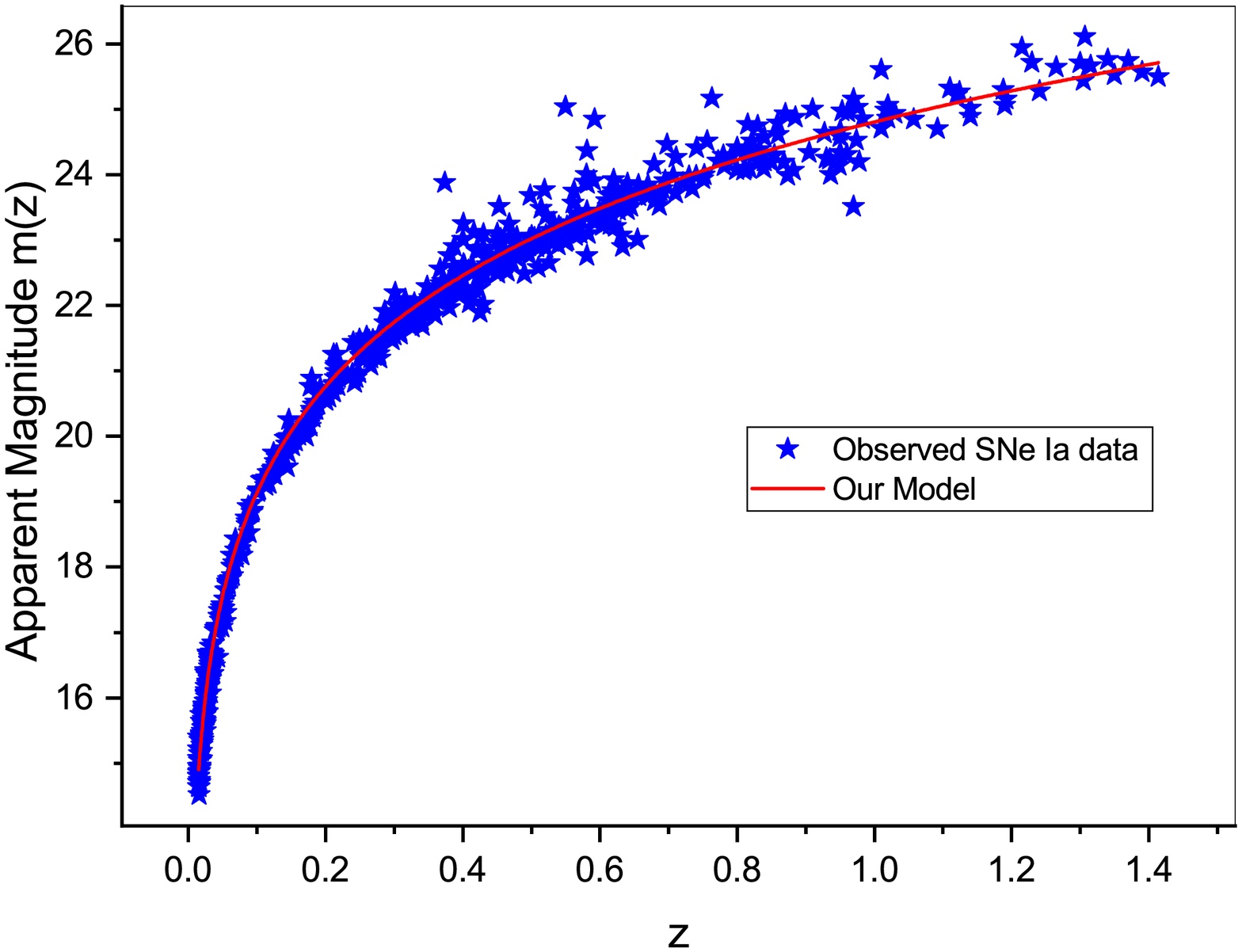}
    b.\includegraphics[width=9cm,height=6cm,angle=0]{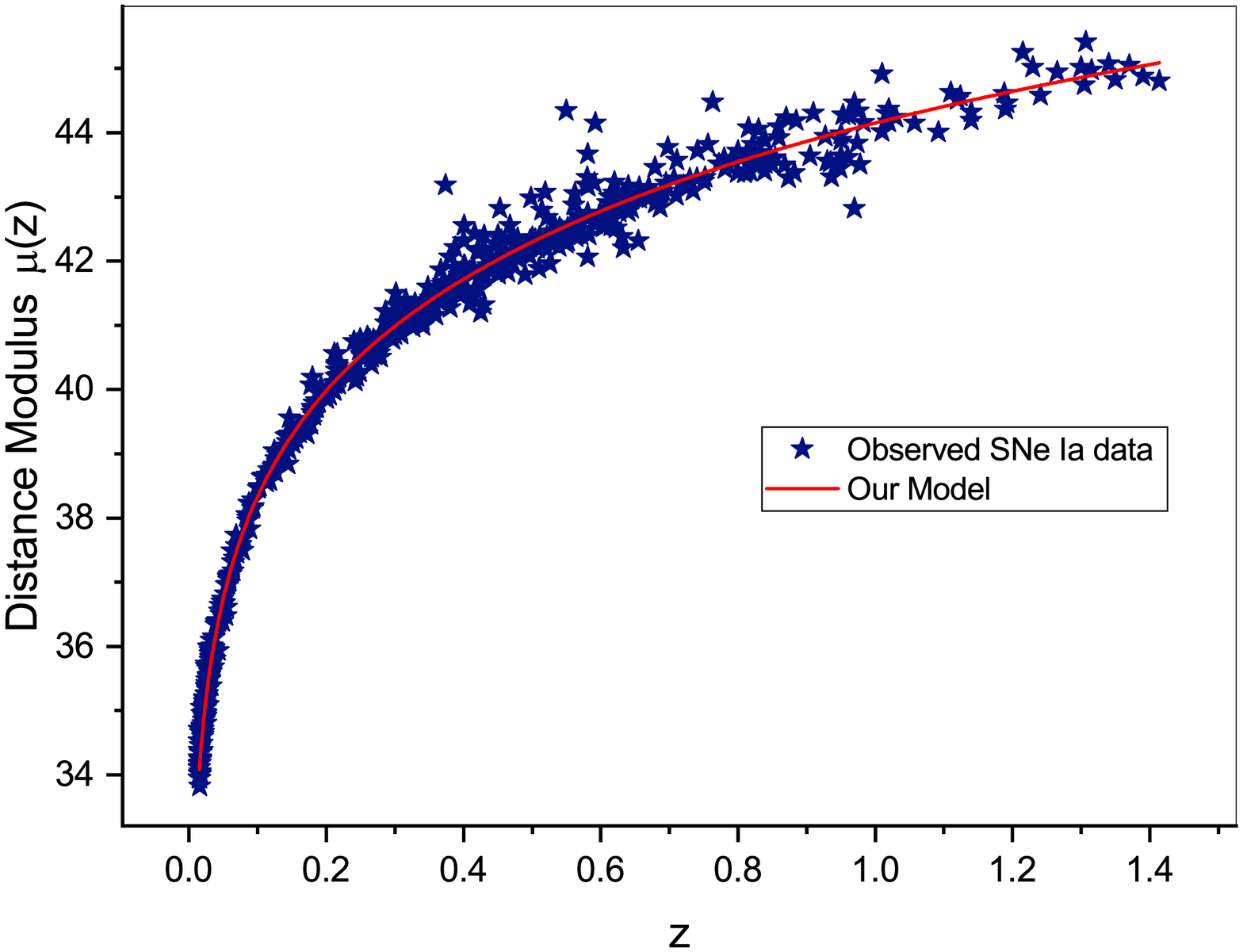}
  	\caption{Best fit curve of apparent magnitude $m(z)$ and distance modulus $\mu(z)$ over redshift $z$ observational data sets respectively.}
\end{figure}
\begin{table}[H]
  \centering
  \begin{tabular}{|c|c|}
     \hline

			Parameter            & Values         \\
			\hline
		
            $\Omega_{m0}$        & $0.44077\pm0.01327$   \\

            $H_{0}$              & $69.7841\pm1.3491$   \\

            $k$                  & $0.34072\pm0.01046$   \\

            $\omega_{\lambda}$   & $0.9342$   \\

            $\chi^{2}$           & $0.06229$   \\
     \hline
   \end{tabular}
  \caption{The best fit values of $\Omega_{m0},~H_{0},~k,~\omega_{\lambda}$ in $\chi^{2}$-test of distance modulus $\mu(z)$ with observed $\mu(z)$ from SNe Ia data with $95\%$ confidence level of bounds. (see Figure 2b ).}\label{T4}
\end{table}
Eqs.~(\ref{eq34}) and (\ref{eq36}) represent the expressions for distance modulus and apparent magnitude $m(z)$ respectively and Figure 2b \& Figure 2a respectively represent their best fit curve over redshift $z$ with observational SNe Ia datasets. The best fit values of $\Omega_{m0},~H_{0},~k,~\omega_{\lambda}$ with SNe Ia datasets are mentioned in Table 3 \& Table 4. The present value of dark energy density parameter is obtained as $\Omega_{\Lambda0}=0.4776$ and corresponding matter and dark energy densities are estimated as $\rho_{0}=6.6058\times10^{-36}~gm/cm^{3}$ and $\Lambda=7.3267\times10^{-36}~gm/cm^{3}$.
\section{Discussion of Results}
\subsection{Cosmographic Analysis}
Recently, cosmography has drawn the most attention of all the rational methods (reference \cite{ref76,ref77}). This model-independent approach \cite{ref78,ref79,ref80} allows the study of dark energy evolution without the requirement to select a particular cosmological model because it just uses the empirical assumptions of the cosmological principle. The typical space flight approach is based on the Taylor expansion of observations that can be directly compared to the data, and the outcomes of such operations are independent of the state equations used to research the evolution of the universe. These factors have made cosmography an effective method for dissecting the cosmological model and have led to its widespread application in attempts to comprehend the kinematics of the universe \cite{ref81}-\cite{ref113}.\\

According to the cosmological principle, the sole degree of freedom that controls the cosmos is a scale factor. In order to determine the cosmographic series coefficients, such as the Hubble parameter $(H)$, deceleration parameter $(q)$, jerk $(j)$, snap $(s)$, lerk $(l)$, and max-out $(m)$ as described in \cite{ref114}, we can expand the current Taylor series of $a(t)$ about the current time.

\begin{equation}\label{eq37}
H=\frac{1}{a}\frac{da}{dt},~~q=-\frac{1}{aH^{2}}\frac{d^{2}a}{dt^{2}},~~j=\frac{1}{aH^{3}}\frac{d^{3}a}{dt^{3}}
\end{equation}
and
\begin{equation}\label{eq38}
s=\frac{1}{aH^{4}}\frac{d^{4}a}{dt^{4}},~~l=\frac{1}{aH^{5}}\frac{d^{5}a}{dt^{5}},~~m=\frac{1}{aH^{6}}\frac{d^{6}a}{dt^{6}}
\end{equation}
These values are used to explore into the dynamics of the late universe. The form of the Hubble expansion may be utilised to identify the properties of the coefficients. The sign of the parameter $q$ in particular indicates whether the universe is expanding or reducing. Positive values of $j$ indicate the occurrence of transitional periods when the universe's expansion varies, and the sign of $j$ determines how the dynamics of the universe change. We also need to know the value of $s$ in order to differentiate between evolving dark energy hypotheses and cosmological constant behaviour.

In this work, we derive these cosmographic coefficients $q,~j,~s$ from the Hubble function $H(z)$ as follows:
\subsection*{Deceleration Parameter}
The deceleration parameter shows the phase of expansion rate of the universe and it is defined in terms of average Hubble parameter $H$ as $q=-\frac{\dot{H}}{H^{2}}-1$. The variation of $q(z)$ over redshift $z$ shows the different phases of the evolution of the universe, $q>0$ shows deceleration, $q=0$ shows phase transition and $q<0$ shows acceleration in expansion. As $z\to0$, we get the present value of $q_{0}$. Thus, the deceleration parameter $q(z)$ is obtained using equation (\ref{eq27}) as follows:
\begin{equation}\label{eq39}
  q(z)=-1+\frac{3\Omega_{m0}(k-k\omega_{\lambda}+2)(1+z)^{3-\frac{3k\omega_{\lambda}}{k+2}}}{2(k+2)\left[\Omega_{m0}(1+z)^{3-\frac{3k\omega_{\lambda}}{k+2}}+\Omega_{\Lambda0}\right]}.
\end{equation}
The expression in Eq.~(\ref{eq39}) represents the deceleration parameter $q(z)$ and its geometrical behaviour is shown in Figure 3. One can see that $q(z)$ is an increasing function of redshift $z$ and $q\to-1$ as $z\to-1$. The present value of $q$ is measured as $q_{0}=-0.4028$ for $H(z)$ datasets and $q_{0}=-0.3692$ for SNe Ia datasets that shows that our universe model is in accelerating phase at present. At $z=0$, $q_{0}$ is obtained as
\begin{equation}\label{eq40}
  q_{0}=-1+\frac{3\Omega_{m0}(k-k\omega_{\lambda}+2)}{2(k+2)\left[\Omega_{m0}+\Omega_{\Lambda0}\right]}.
\end{equation}
From which, we can obtained the condition for acceleration in expansion $q<0$ as
\begin{equation}\label{eq41}
  \Omega_{\Lambda0}>\frac{(2k+1)(k-3k\omega_{\lambda}+2)}{(k+2)^{2}(k-k\omega_{\lambda}+2)}
\end{equation}
For $k=1$, $\omega_{\lambda}$ vanishes and model represents a flat $\Lambda$CDM model as $q_{0}=\frac{1}{2}(1-3\Omega_{\Lambda})$ that reveals the condition for acceleration $q<0$ when $\Omega_{\Lambda}>\frac{1}{3}$ and as $\Omega_{\Lambda}\to1$, $q\to-1$. Also, from Figure 3a, we can see that $q$ shows a signature-flipping (decelerating to accelerating) and hence, we have calculated the transition redshift $z_{t}$ such that $q(z_{t})=0$ which is obtained as follows:
 \begin{equation}\label{eq42}
  z_{t}=\left[\frac{2\Omega_{\Lambda0}}{\Omega_{m0}}\frac{k+2}{k-3k\omega_{\lambda}+2}\right]^{\frac{k+2}{3(k-k\omega_{\lambda}+2)}}-1
\end{equation}
 SNe type Ia measurements offer the most conclusive empirical proof of the change from past deceleration to present acceleration. The SNe data support recent acceleration ($z<0.5$) and prior deceleration ($z > 0.5$), according to their preliminary study. Recent results from the High-z Supernova Search (HZSNS) team include $z_{t}=0.46\pm0.13$ at $(1; \sigma)$ c.l. \cite{ref2,ref3,ref115} in 2004 and $z_{t}=0.43\pm0.07$ at $(1; \sigma)$ c.l. \cite{ref115} in 2007. The transition redshift obtained from the Supernova Legacy Survey (SNLS) by Astier et al. \cite{ref116} and Davis et al. \cite{ref69} is $z_{t}\sim 0.6$ $(1; \sigma)$, which is more in line with the flat $\Lambda$CDM model ($z_{t} = (2\Omega_{\Lambda}/\Omega_{m})^{\frac{1}{3}} - 1 \sim 0.66$). An additional restriction is $0.60 \leq z_{t}\leq 1.18$ ($2 \sigma$, joint analysis) \cite{ref117}. Additionally, the transition redshift for our deduced model is $z_{t}\cong 0.669,~0.576$ along two datasets (see Figure 3a), which is in good accord with the Type Ia supernovae Hubble diagram, including the farthest known supernova SNI997ff at $z\approx 1.7$ (Riess et al.; Amendola; \cite{ref118}). This discovery indirectly demonstrates that the transition redshift can be used to reevaluate the history of cosmic expansion.
\begin{figure}[H]
\centering
	\includegraphics[width=10cm,height=8cm,angle=0]{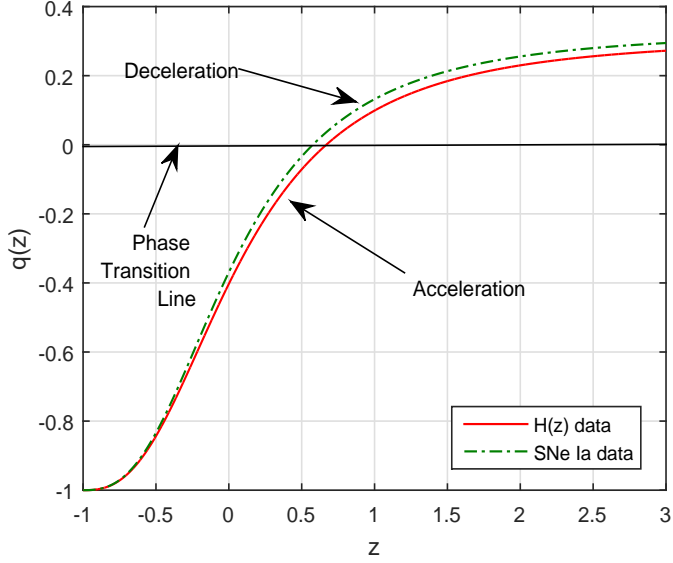}
   	\caption{The behaviour of deceleration parameter $q(z)$ over redshift $z$.}
\end{figure}
Equation (\ref{eq42}) represents the expression for transition redshift in our model and one can see that $z_{t}\to (2\Omega_{\Lambda}/\Omega_{m})^{\frac{1}{3}} - 1$ as $(k, \omega_{\lambda})\to(1, 0)$ that shows $\Lambda$CDM value.
\subsection*{Jerk Parameter}
The third cosmographic coefficient jerk parameter is obtained as
\begin{eqnarray*}
  j(z)&=&1-\frac{3\Omega_{m0}(k-k\omega_{\lambda}+2)(1+z)^{3-\frac{3k\omega_{\lambda}}{k+2}}}{(k+2)\left[\Omega_{m0}(1+z)^{3-\frac{3k\omega_{\lambda}}{k+2}}+\Omega_{\Lambda0}\right]}+\frac{9\Omega_{m0}^{2}(k-k\omega_{\lambda}+2)^{2}(1+z)^{6-\frac{6k\omega_{\lambda}}{k+2}}}{4(k+2)^{2}\left[\Omega_{m0}(1+z)^{3-\frac{3k\omega_{\lambda}}{k+2}}+\Omega_{\Lambda0}\right]^{2}}\\
  &&+\frac{\Omega_{m0}(3k\omega_{\lambda}-2k-4)(3k\omega_{\lambda}-3k-6)(1+z)^{2-\frac{3k\omega_{\lambda}}{k+2}}}{2(k+2)^{2}\left[\Omega_{m0}(1+z)^{3-\frac{3k\omega_{\lambda}}{k+2}}+\Omega_{\Lambda0}\right]}-\frac{\Omega_{m0}^{2}(3k\omega_{\lambda}-3k-6)^{2}(1+z)^{5-\frac{6k\omega_{\lambda}}{k+2}}}{4(k+2)^{2}\left[\Omega_{m0}(1+z)^{3-\frac{3k\omega_{\lambda}}{k+2}}+\Omega_{\Lambda0}\right]^{2}}.
\end{eqnarray*}
\begin{equation}\label{eq43}
\end{equation}
\begin{figure}[H]
	a.\includegraphics[width=9cm,height=8cm,angle=0]{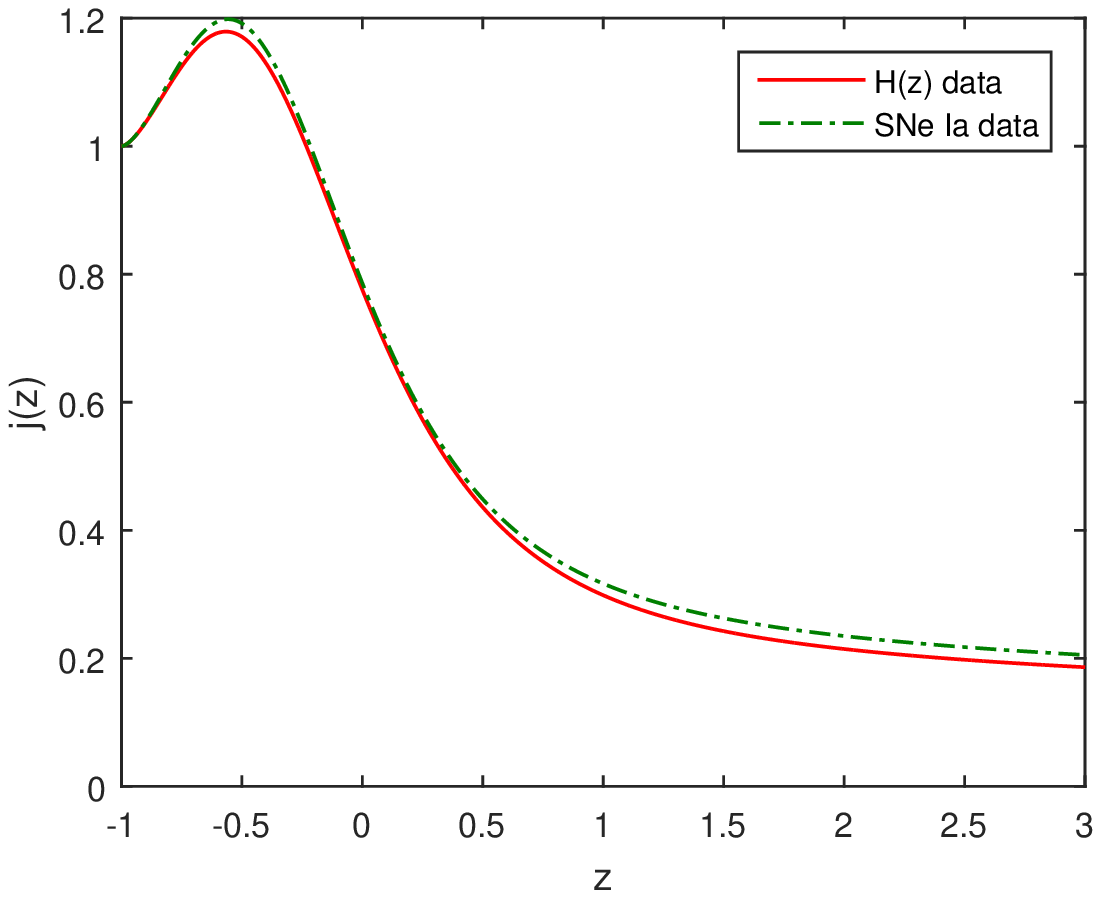}
    b.\includegraphics[width=9cm,height=8cm,angle=0]{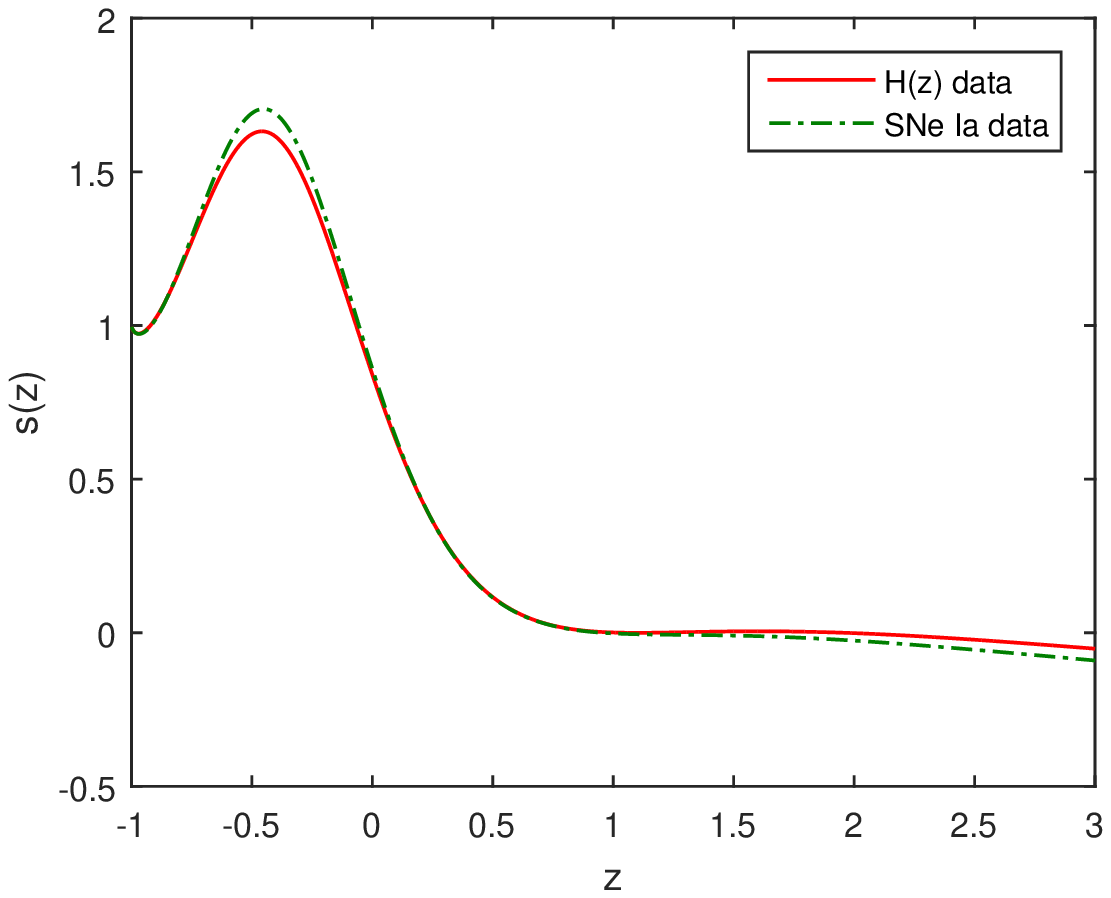}
  	\caption{The variation of jerk $j(z)$ and snap $s(z)$ parameter over redshift $z$ respectively.}
\end{figure}
Equation (\ref{eq43}) shows the expression for jerk parameter $j(z)$ and Figure 4a reveals its evolution with redshift $z$. One can see that it varies over $(0.2, 1.4)$ and $j\to1$ as $z\to-1$. Also, we can see that $j>0$ over the range $(0\leq z \leq 3)$ that reveals the transition phase in evolution of the universe \cite{ref2,ref3,ref69} and \cite{ref115}-\cite{ref118}. The present value of jerk parameter is estimated as $j_{0}=0.7758,~0.7861$ along two datasets respectively $H(z)$ datasets and SNe Ia datasets.
\subsection*{Snap Parameter}
The fourth cosmographic coefficient snap parameter $s(z)$ in our model is obtained as
\begin{eqnarray*}
  s(z)
  &=&1-\frac{9\Omega_{m0}(k-k\omega_{\lambda}+2)}{2(k+2)\left[\Omega_{m0}+\Omega_{\Lambda0}(1+z)^{\frac{3k\omega_{\lambda}}{k+2}-3}\right]}+\frac{27\Omega_{m0}^{2}(k-k\omega_{\lambda}+2)^{2}}{4(k+2)^{2}\left[\Omega_{m0}+\Omega_{\Lambda0}(1+z)^{\frac{3k\omega_{\lambda}}{k+2}-3}\right]^{2}}\\
  & &-\frac{27\Omega_{m0}^{3}(k-k\omega_{\lambda}+2)^{3}}{8(k+2)^{3}\left[\Omega_{m0}+\Omega_{\Lambda0}(1+z)^{\frac{3k\omega_{\lambda}}{k+2}-3}\right]^{3}}+\frac{3\Omega_{m0}(3k\omega_{\lambda}-2k-4)(3k\omega_{\lambda}-3k-6)(1+z)^{2-\frac{3k\omega_{\lambda}}{k+2}}}{2(k+2)^{2}\left[\Omega_{m0}(1+z)^{3-\frac{3k\omega_{\lambda}}{k+2}}+\Omega_{\Lambda0}\right]}\\
  &&-\frac{3\Omega_{m0}^{2}(3k\omega_{\lambda}-3k-6)^{2}(1+z)^{5-\frac{6k\omega_{\lambda}}{k+2}}}{4(k+2)^{2}\left[\Omega_{m0}(1+z)^{3-\frac{3k\omega_{\lambda}}{k+2}}+\Omega_{\Lambda0}\right]^{2}}-\frac{\Omega_{m0}(3k\omega_{\lambda}-2k-4)(3k\omega_{\lambda}-3k-6)(1+z)^{3-\frac{3k\omega_{\lambda}}{k+2}}}{2(k+2)^{2}\left[\Omega_{m0}(1+z)^{3-\frac{3k\omega_{\lambda}}{k+2}}+\Omega_{\Lambda0}\right]}\\
  &&+\frac{\Omega_{m0}^{2}(3k\omega_{\lambda}-3k-6)^{2}(1+z)^{6-\frac{6k\omega_{\lambda}}{k+2}}}{4(k+2)^{2}\left[\Omega_{m0}(1+z)^{3-\frac{3k\omega_{\lambda}}{k+2}}+\Omega_{\Lambda0}\right]^{2}}+\frac{3\Omega_{m0}^{3}(3k\omega_{\lambda}-3k-6)^{3}(1+z)^{7-\frac{9k\omega_{\lambda}}{k+2}}}{8(k+2)^{3}\left[\Omega_{m0}(1+z)^{3-\frac{3k\omega_{\lambda}}{k+2}}+\Omega_{\Lambda0}\right]^{3}}\\
  &&-\frac{3\Omega_{m0}^{2}(3k\omega_{\lambda}-3k-6)^{2}(3k\omega_{\lambda}-2k-4)(1+z)^{4-\frac{3k\omega_{\lambda}}{k+2}}}{4(k+2)^{3}\left[\Omega_{m0}(1+z)^{3-\frac{3k\omega_{\lambda}}{k+2}}+\Omega_{\Lambda0}\right]^{2}}\\
  &&+\frac{\Omega_{0}(3k\omega_{\lambda}-k-2)(3k\omega_{\lambda}-2k-4)(3k\omega_{\lambda}-3k-6)(1+z)^{1-\frac{3k\omega_{\lambda}}{k+2}}}{2(k+2)^{3}\left[\Omega_{m0}(1+z)^{3-\frac{3k\omega_{\lambda}}{k+2}}+\Omega_{\Lambda0}\right]^{\frac{1}{2}}}\\
  &&-\frac{9\Omega_{m0}^{3}(k-k\omega_{\lambda}+2)^{3}(2+z)(1+z)^{8-\frac{9k\omega_{\lambda}}{k+2}}}{4(k+2)^{3}\left[\Omega_{m0}(1+z)^{3-\frac{3k\omega_{\lambda}}{k+2}}+\Omega_{\Lambda0}\right]^{3}}-\frac{9\Omega_{m0}^{2}\Omega_{\Lambda0}(k-k\omega_{\lambda}+2)^{2}(2k-3k\omega_{\lambda}+4)(2+z)(1+z)^{5-\frac{6k\omega_{\lambda}}{k+2}}}{2(k+2)^{3}\left[\Omega_{m0}(1+z)^{3-\frac{3k\omega_{\lambda}}{k+2}}+\Omega_{\Lambda0}\right]^{3}}.
  \end{eqnarray*}
\begin{equation}\label{eq44}
\end{equation}
Equation (\ref{eq44}) represents the expression for snap parameter $s(z)$ and its geometrical behaviour is shown in Figure 4b. We can see that $s(z)$ varies as $(-0.5, 2)$ over the range $(0\leq z \leq 3)$ and as $z\to-1$, $s\to1$ that shows the existence of dark energy in the universe \cite{ref81}-\cite{ref113}. The present value of snap parameter is estimated as $s_{0}=0.8398,~0.8594$ along two datasets respectively $H(z)$ datasets and SNe Ia datasets.\\

Recently Capozziello et al. \cite{ref114} have studies extended gravity theories using  cosmography in details and in \cite{ref119}-\cite{ref122}, they have used cosmography to reconstruct various cosmological models in different modified gravity theories. The values of $H_{0},~q_{0},~j_{0}$ and $s_{0}$ obtained in our model are very closed to used values in \cite{ref114} and \cite{ref119}-\cite{ref122}.
\subsection{Energy Density}
\begin{figure}[H]
	a.\includegraphics[width=9cm,height=8cm,angle=0]{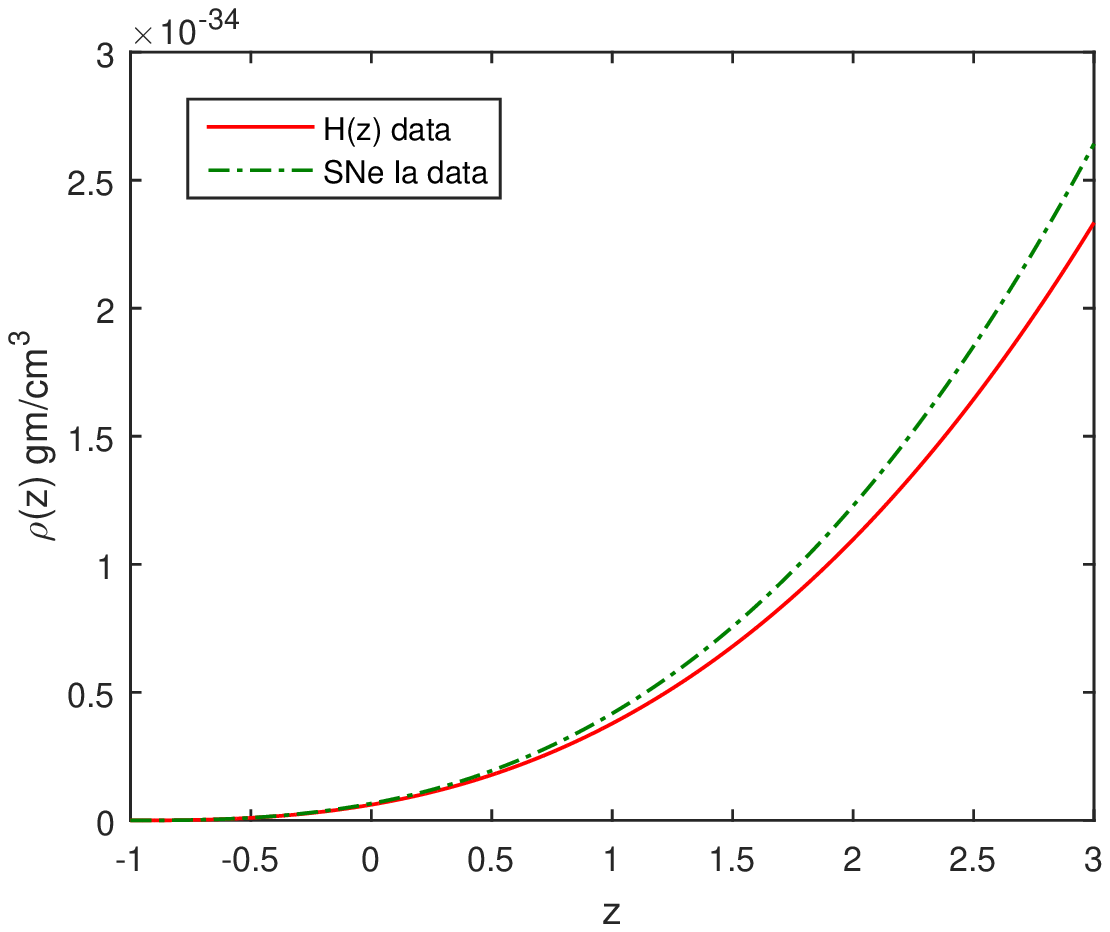}
    b.\includegraphics[width=9cm,height=8cm,angle=0]{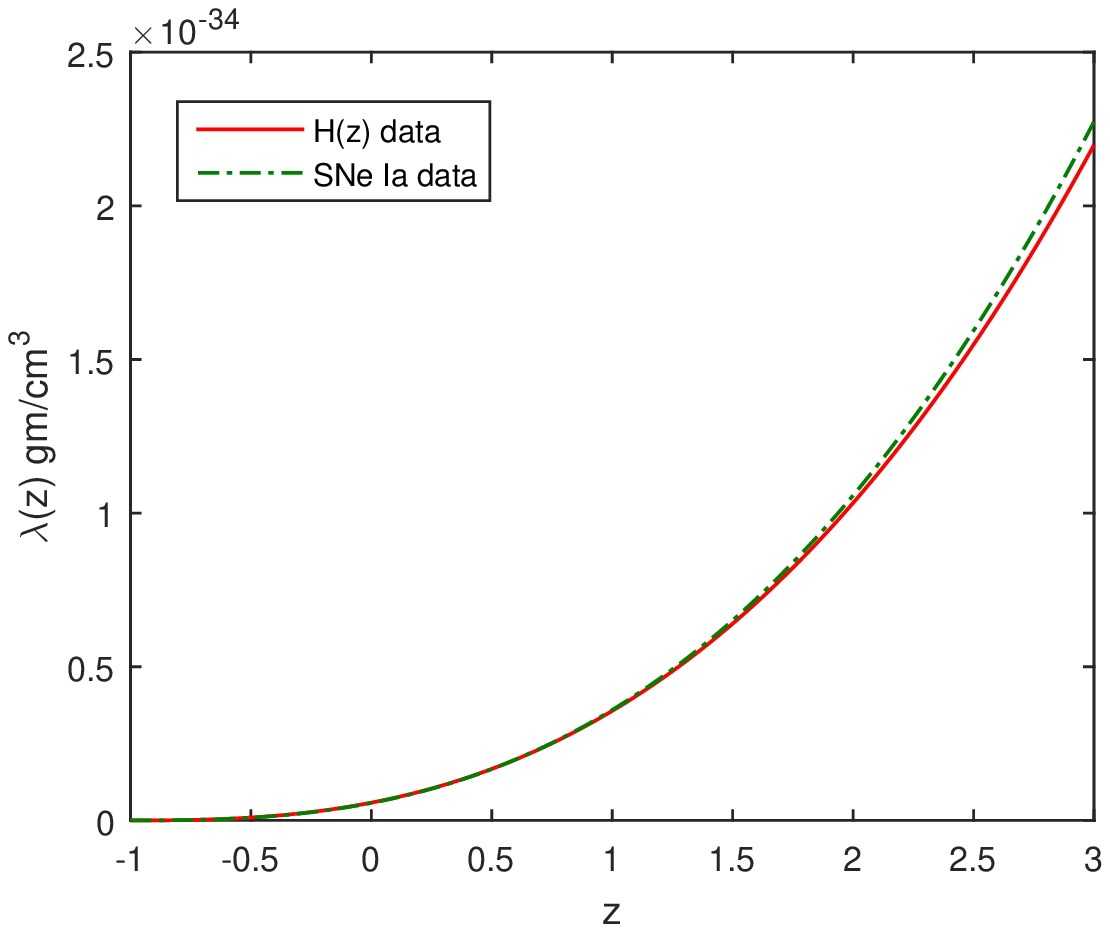}
  	\caption{The behaviour of total energy density $\rho(z)$ and string tension density $\lambda(z)$ over $z$ respectively.}
\end{figure}
Eqs.~(\ref{eq23}) \& (\ref{eq28}) represent the expressions for total energy density $\rho(z)$ and string tension density $\lambda(z)$ respectively and their geometrical behaviour are shown in Figure 5a \& Figure 5b. One can see that the total energy density $\rho$ is increasing with redshift $z$ and as $z\to-1$, $\rho\to0$ and as $z\to\infty$, $\rho\to\infty$. One can see that $\lambda(z)$ is an increasing function of $z$ which shows that early universe is string dominated and with evolution of the universe it decreases to zero. From the Table 2, one can see that $\omega_{\lambda}>0$ which shows that $\lambda>0$ as given in \cite{ref51,ref52}. Reduction in string tension density causes the acceleration in expansion of the universe. We can see that the particle energy density $\rho_{p}>0$ or $\rho-\lambda>0$ i.e. all energy conditions are satisfied.
\begin{figure}[H]
\centering
	\includegraphics[width=10cm,height=8cm,angle=0]{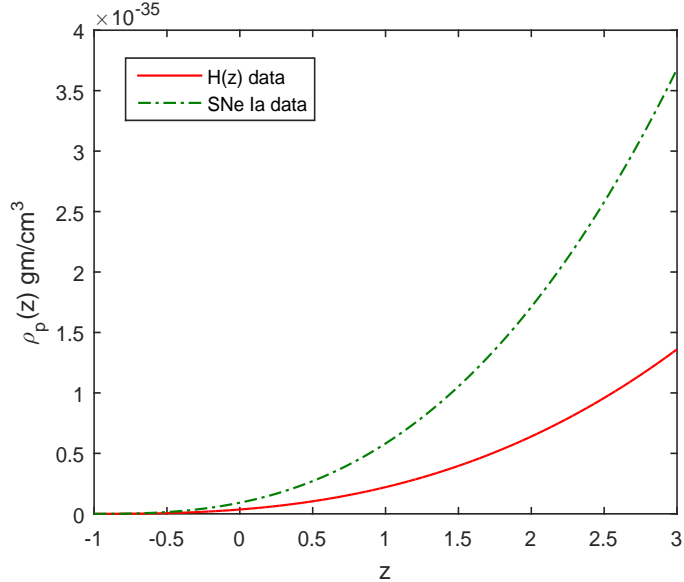}
   	\caption{The variation of particle energy density $\rho_{p}$ with redshift $z$.}
\end{figure}
The energy density parameters $\Omega_{m}$ and $\Omega_{\Lambda}$ are given by Eq.~(\ref{eq25}) and their geometrical behaviour are shown in the figures 7a \& 7b. One can see that matter energy density parameter $\Omega_{m}$ is an increasing function of $z$ that shows early universe is matter dominated and the dark energy density parameter $\Omega_{\Lambda}$ is a decreasing function of redshift $z$ and $\Omega_{\Lambda}\to1$ as $z$ tends to minus one that shows late-time universe is dark energy dominated. From the Table 3, one can see that the total energy density parameter $\Omega<1$ that shows our universe model is closed universe and it tends to flat geometry $\Omega\to1$ that shows late-time universe tends to $\Lambda$CDM model.
\begin{figure}[H]
	a.\includegraphics[width=9cm,height=8cm,angle=0]{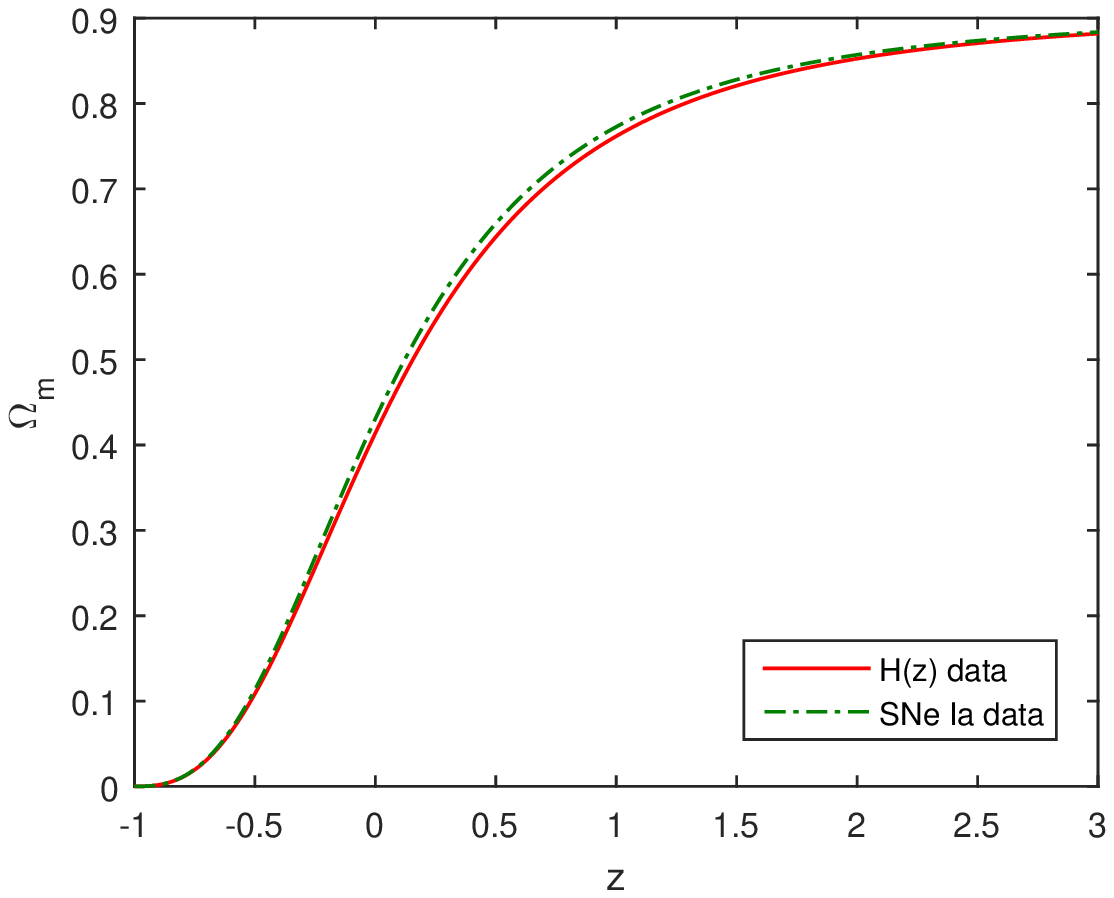}
    b.\includegraphics[width=9cm,height=8cm,angle=0]{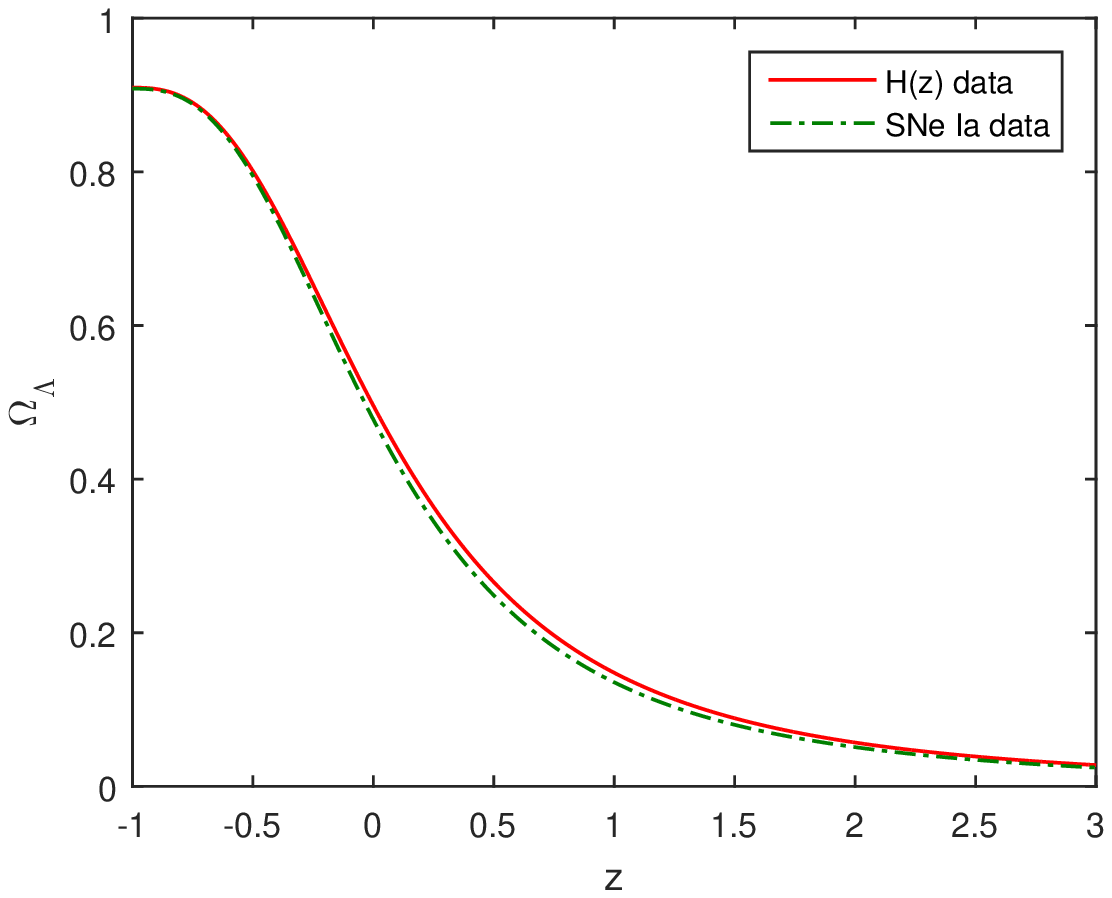}
  	\caption{The variation of energy density parameters $\Omega_{m}$ and $\Omega_{\Lambda}$ with redshift $z$ respectively.}
\end{figure}
\subsection{Om Diagnostic Analysis}
The cosmic dark energy models can be classified through behaviour of Om diagnostic function \cite{ref70}. The diagnostic for a spatially homogeneous is given by

\begin{equation}\label{eq45}
  Om(z)=\frac{\left(\frac{H(z)}{H_{0}}\right)^{2}-1}{(1+z)^{3}-1},
\end{equation}
where $H(z)$ is the average Hubble parameter given in Eq.~(\ref{eq27}) and $H_{0}$ is its current value. A negative slope of $Om(z)$ corresponds to quintessence motion, and a positive slope corresponds to phantom motion. The $Om(z)$ constant represents the $\Lambda$CDM model.\\
Figure 8 shows the geometrical behaviour of Om diagnostic function $Om(z)$ over redshift $z$ for the model and mathematical expression is given in above equation (\ref{eq45}). From figure 8, one can see that the slope of $Om(z)$ function is negative for that shows quintessence behaviour of the model and at late-time as $z\to-1$, it tends to a constant value that shows the tendency of model to $\Lambda$CDM model at late-time universe. Thus, we found that our model is quintessence dark energy model.
\begin{figure}[H]
\centering
	\includegraphics[width=10cm,height=9cm,angle=0]{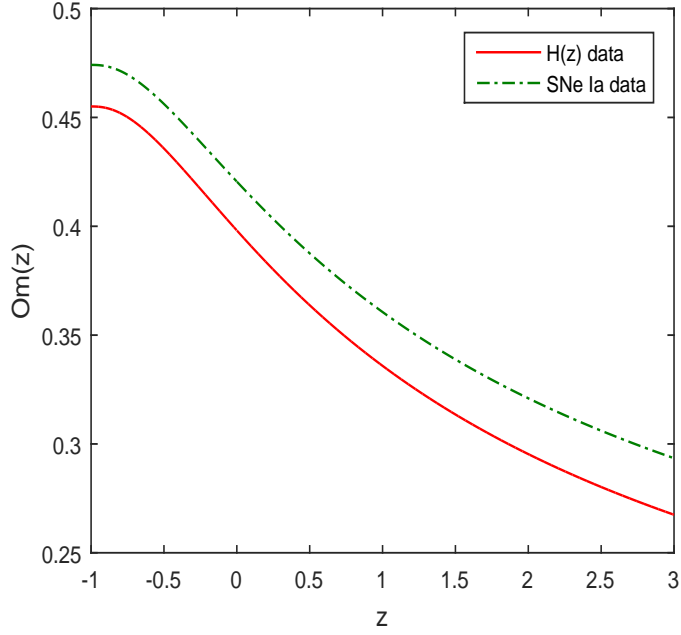}
  	\caption{The variation of Om diagnostic function $Om(z)$ over redshift $z$.}
\end{figure}

\section{Conclusions}
      In this paper, we have investigated an anisotropic cosmological model in $f(Q)$ gravity with string fluid in LRS Bianchi type-I universe. We have considered the arbitrary function $f(Q)=Q+\alpha\sqrt{~Q}+2\Lambda$ where $\alpha$ is model free parameter and $\Lambda$ is the cosmological constant. We have established a relationship between matter energy density $\Omega_{m}$ and dark energy density $\Omega_{\Lambda}$ through Hubble function using using constant equation of state parameter $\omega, \omega_{\lambda}$ with $p=\omega\rho$ \& $\lambda=\omega_{\lambda}\rho$. We have made observational constraint on the model using $\chi^2$-test with $46$ observed Hubble datasets $H(z)$ and $715$ observed datasets of SNe Ia and obtained the best fit values of cosmological parameters. We have used these best fit values to obtain the result and discussion. The estimated values of various parameters in our model is given in the following Table 6:
\begin{table}[H]
  \centering
  \begin{tabular}{|c|c|c|}
     \hline

			Parameters             & For $H(z)$ data               & SNe Ia data  \\
			\hline
			$k$                    & $0.30652$                 & $0.30242$      \\
            $\Omega$               & $0.90962$              & $0.90821$      \\
            $\Omega_{m0}$          & $0.41392$             & $0.43061$      \\
            $\Omega_{\Lambda0}$    & $0.4957$             & $0.4776$      \\
            $\omega_{\lambda}$     & $0.9417$             & $0.8607$      \\
            $H_{0}$                & $68.6095$            & $69.7841$      \\
            $\rho_{0}$             & $6.1378\times10^{-36}$& $6.6058\times10^{-36}$      \\
            $\Lambda$              & $7.3505\times10^{-36}$& $7.3267\times10^{-36}$      \\
            $\alpha$               & $1462.0$              & $2414.9$      \\
            $q_{0}$                & $-0.4028$             & $-0.3692$      \\
            $j_{0}$                & $0.7758$              & $0.7861$      \\
            $s_{0}$                & $0.8398$              & $0.8594$      \\
            $z_{t}$                & $0.669$              & $0.576$      \\
        \hline
   \end{tabular}
  \caption{The estimated values of various cosmological parameters in our model.}\label{T5}
\end{table}
 The main features of our derived model are as follows:
 \begin{itemize}
   \item The present model shows a signature-flipping (decelerating to accelerating) point at $z_{t}=0.669,~0.576$ respectively for two datasets which are good features of our model and it is supported by recent observations \cite{ref2,ref3,ref69} and \cite{ref115}-\cite{ref118}.
   \item We have found that the early universe is anisotropic and string dominated and decreasing string tension density may causes the acceleration in expansion.
   \item The total energy density parameter $\Omega<1$ and hence, the derived model is a closed universe model and in present $\Omega_{m}<\Omega_{\Lambda}$ that shows that the present universe is dark energy dominated.
   \item We have found the present value of cosmographic coefficients $H_{0}=68.6095,~69.7841~ Km/s/Mpc$, deceleration parameter $q_{0}<0$ and jerk parameter $j_{0}>0$ and so, the present phase of our universe model is accelerating. The positive signature of jerk parameter $j>0$ supported the phase transition in our universe model. The values snap parameter $(-0.5, 2)$ supported dark energy dominated universe \cite{ref119}-\cite{ref122}.
   \item We have found that all the energy conditions are satisfied and we have estimated the present values of energy densities $\rho_{0}$ and $\rho_{p}$ (see Table 5).
   \item We have analysed the Om diagnostic analysis for anisotropic universe and found that our universe is quintessence dark energy dominated universe model.
   \item We have estimated the value of model free parameter $\alpha=1462.0,~2414.9$ respectively for two datasets $H(z)$ and SNe Ia data.
 \end{itemize}
\section*{Acknowledgement}
A. Pradhan \& A. Dixit thank the IUCAA, Pune, India for providing facilities under associateship programs. D.C. Maurya is thankful to IASE (Deemed to be University), GVM, Sardarshahr, Rajasthan, India to providing facilities and support where part of this work is carried out.  The authors are thankful to the anonymous reviewers for their valuable suggestions and comments for the improvement of the paper.

\end{document}